\documentclass[shortnote,twocolumn,amsmath,amssymb,amsbsy]{jpsj3}
\usepackage{txfonts}
\usepackage[dvipdfmx]{graphicx}
\usepackage{bm}
\usepackage{braket}
\usepackage{mathtools}
\usepackage{color}

\title{On equivalence 
of two formulas of orbital magnetic susceptibility
for tight-binding models}

\author{Tomonari Mizoguchi$^1$\thanks{mizoguchi@rhodia.ph.tsukuba.ac.jp} and Nobuyuki Okuma$^2$}
\inst{$^1$Department of Physics, University of Tsukuba, Tsukuba, Ibaraki 305-8571, Japan\\
$^2$ Graduate School of Engineering, Kyushu Institute of Technology, Kitakyushu 804-8550, Japan }

\abst{We prove that 
two formulas of the orbital magnetic susceptibility,
namely, Koshino-Ando's formula and G\'{o}mez-Santos and Stauber's formula,
are equivalent for the tight-binding models. 
The difference between these two formulas can be 
written in a form containing the 
surface terms arising from the momentum-space integration,
and we show that the surface terms are exactly vanishing
although the primitive functions are 
seemingly non-periodic 
with respect to 
the translation 
by the reciprocal lattice vector. 
}

\begin{document}
\maketitle
Magnetic-field effects in solids have long  drawn interest of condensed matter physicists. 
Orbital magnetic susceptibility is 
a fundamental quantity that describes the magnetic-field response 
of nonmagnetic materials coming from orbital motion of electrons.
Even for the noninteracting electrons,
the theoretical treatment of magnetic field is challenging
and a tremendous amount of effort has been devoted~\cite{Landau1930,Peierls1933,Hebborn1959,McClure1960,Blount1962,Fukuyama1969,Fukuyama1971,Koshino2007_2,GomezSantos2011,Raoux2015,Ogata2015,Gao2015,Piechon2016}. 
Through these works, 
it has been emphasized that the interband effect
plays an essential role in the orbital magnetic susceptibility, and some parts of the interband effects are reinterpreted in terms of modern notions 
such as topology and quantum geometry.
Such an aspect also gains attention from the experimental point of view~\cite{Vallejo2021}.

The aim of this Short Note 
is to show the equivalence of 
two theoretical formulas 
for the orbital magnetic susceptibility 
in a detailed manner.
One formula is given 
by Koshino and Ando~\cite{Koshino2007_2}:
\begin{align}
\chi^{\rm (KA)}_{\rm orb} =  
-\frac{e^2}{4 \pi \hbar^2}   
\int_{-\infty}^{\infty} d\epsilon f(\epsilon)  
\mathrm{Im} [\Theta^{\mathrm{(KA)}}(\epsilon)],
\end{align}
with 
\begin{align} 
 \Theta(\epsilon)^{\rm (KA)} =&
 \frac{1}{V}\sum_{\bm{k}}  
\mathrm{Tr} 
\{v^x_{\bm{k}} G_{\bm{k}}(\epsilon)v^y_{\bm{k}} G_{\bm{k}}(\epsilon)v^x_{\bm{k}} G_{\bm{k}}  (\epsilon)v^y_{\bm{k}} G_{\bm{k}}(\epsilon)\notag \\
-& 2 v^x_{\bm{k}} G_{\bm{k}}(\epsilon)v^x_{\bm{k}} G_{\bm{k}}(\epsilon)v^y_{\bm{k}} G_{\bm{k}}  (\epsilon)v^y_{\bm{k}} G_{\bm{k}}(\epsilon) \notag \\
-& \frac{1}{2} 
[G_{\bm{k}}(\epsilon)v^y_{\bm{k}} G_{\bm{k}}(\epsilon)\frac{\partial v^x_{\bm{k}}}{\partial k_x}
G_{\bm{k}}(\epsilon)v^y_{\bm{k}}
\notag \\
+& G_{\bm{k}}(\epsilon)v^x_{\bm{k}} G_{\bm{k}}(\epsilon)\frac{\partial v^y_{\bm{k}}}{\partial k_y}
G_{\bm{k}}(\epsilon)v^x_{\bm{k}}]
\}.
\end{align}
The other one is given by G\'{o}mez-Santos and Stauber~\cite{GomezSantos2011}:
\begin{align}
\chi^{\rm (GS)}_{\rm orb} = 
-\frac{e^2}{4\pi \hbar^2}   
\int_{-\infty}^{\infty} d\epsilon f(\epsilon)  \mathrm{Im} [\Theta^{\rm (GS)}(\epsilon)],
\end{align}
with 
\begin{align} 
 \Theta(\epsilon)^{\rm (GS)} = 
 \frac{1}{V}\sum_{\bm{k}}  
\mathrm{Tr} 
[2 v^x_{\bm{k}} G_{\bm{k}}(\epsilon)v^y_{\bm{k}} G_{\bm{k}}(\epsilon)v^x_{\bm{k}} G_{\bm{k}}  (\epsilon)v^y_{\bm{k}} G_{\bm{k}}(\epsilon)\notag \\
+ \left( G_{\bm{k}}(\epsilon)v^x_{\bm{k}} G_{\bm{k}}(\epsilon)v^y_{\bm{k}} + G_{\bm{k}}(\epsilon)v^y_{\bm{k}} G_{\bm{k}}(\epsilon) v^x_{\bm{k}}\right) G_{\bm{k}}  (\epsilon)\frac{\partial v^y_{\bm{k}}}{\partial {k_x}} ].
\end{align}
Here, 
$e$ is the elementary charge, $\hbar$ is the reduced Planck constant, $V$ is the volume of the system, 
and 
$f(\epsilon) = 1/(e^{\beta(\epsilon-\mu)}+1)$ is the Fermi-Dirac distribution function.
$v^\mu_{\bm{k}}$ ($\mu  =x, y$) and 
$G_{\bm{k}}(\epsilon)$
are the matrices of the velocity and the Green's  function, respectively, whose precise definitions are presented later. 

It has been pointed out in Ref.~\cite{Ogata2016_2} 
that one can show the equivalence 
of the above two formulas by
using the partial integration 
for the momentum-space integration.
However, the equivalence holds only up to the surface term~\cite{remark}.
For concreteness and the simplicity of writing, in the following, we consider the case of two-dimensional systems having a square unit cell with the lattice constant being unity, 
but the argument for the generic tight-binding models (in both two and three dimensions) is 
essentially the same. 
In this case, we have 
\begin{align}
\Theta(\epsilon)^{\rm (KA)}
-\Theta(\epsilon)^{\rm (GS)}
&\propto  
\int_{0}^{2\pi} 
dk_y [\mathrm{Tr}\{ 
G_{\bm{k}}(\epsilon)v^y_{\bm{k}}
G_{\bm{k}}(\epsilon)v^x_{\bm{k}}
G_{\bm{k}}(\epsilon)v^y_{\bm{k}}
\}]_{k_x = 0}^{2\pi}  \notag \\
&+ \int_{0}^{2\pi} d k_x [
\mathrm{Tr}\{
G_{\bm{k}}(\epsilon)v^x_{\bm{k}}
G_{\bm{k}}(\epsilon)v^y_{\bm{k}}
G_{\bm{k}}(\epsilon)v^x_{\bm{k}}
\}
]_{k_y=0}^{2\pi}. \label{eq:GSvsKA}
\end{align}
To obtain the above, the relation $\frac{\partial G_{\bm{k}}(\epsilon)}{\partial k_{\mu}} = G_{\bm{k}}(\epsilon) 
v^\mu_{\bm{k}}G_{\bm{k}}(\epsilon)$
as well as the cyclic property of the trace have been used. 
Now, let us clarify the issue we want to discuss in this note.
Some of the readers may think
that the right-hand side of Eq.~(\ref{eq:GSvsKA}) is obviously vanishing
since the primitive functions,
$\mathrm{Tr}\{ 
G_{\bm{k}}(\epsilon)v^y_{\bm{k}}
G_{\bm{k}}(\epsilon)v^x_{\bm{k}}
G_{\bm{k}}(\epsilon)v^y_{\bm{k}}
\}$ and $\mathrm{Tr}\{
G_{\bm{k}}(\epsilon)v^x_{\bm{k}}
G_{\bm{k}}(\epsilon)v^y_{\bm{k}}
G_{\bm{k}}(\epsilon)v^x_{\bm{k}}
\}$,
have periodicity of $k_{\mu} 
\rightarrow k_{\mu} + 2\pi$.
However, 
the periodicity is nontrivial 
when there are multiple number of sublattices in the unit cell.
In such a case, the formula is valid when a Fourier transform is performed with incorporating the sublattice position, as we will elaborate later.
Then, at this stage, it is not clear 
whether Eq.~(\ref{eq:GSvsKA}) is vanishing.
What we shall show in this note is 
that this periodicity indeed holds 
even under
the sublattice-position-dependent Fourier transformation. 
This leads to the conclusion that 
the two formulas are equivalent 
as far as the tight-binding models are concerned~\cite{remark2}. 

Let us first manifest 
the set-up of the present study. 
We consider the tight-binding Hamiltonian 
which is generically written as
\begin{align}
H = \sum_{\bm{R}, \bm{R}^\prime} \sum_{a,b} h^{ab}(\bm{R} - \bm{R}^\prime) 
c^{\dagger}_{\bm{R},a} c_{\bm{R}^\prime,b}, \label{eq:Ham_TB}
\end{align}
where $\bm{R}$, $\bm{R}^\prime$ denote the position of the center of the unit cells, and $a$, $b$ label 
the internal degrees of freedom including 
the sublattice degrees of freedom.
The translational invariance lies in the fact that $h^{ab}(\bm{R}- \bm{R}^\prime)$ is a function of $\bm{R}-\bm{R}^\prime$.

We perform 
a Fourier 
transform
with incorporating 
the position dependence 
of each sublattice inside the unit cell as
$c_{\bm{k},a} =\frac{1}{\sqrt{N}} \sum_{\bm{R}} c_{\bm{R},a} e^{-i \bm{k} \cdot (\bm{R} + \bm{r}_a )}$ and  
$c^\dagger_{\bm{k},a} =\frac{1}{\sqrt{N}} \sum_{\bm{R}} c^\dagger_{\bm{R},a} e^{i \bm{k} \cdot (\bm{R} + \bm{r}_a)}$, 
where $N$ denotes the number of the unit cells 
and $\bm{r}_a$ denotes the position of the sublattice $a$ 
measured from the center of the unit cell. 
Using this convention, 
the Hamiltonian 
of Eq.~(\ref{eq:Ham_TB}) is rewritten as
$H = \sum_{\bm{k}} \sum_{a,b} h_{\bm{k}}^{ab} c^{\dagger}_{\bm{k},a}c_{\bm{k},b}$,
where
$h_{\bm{k}}^{ab}  = \sum_{\bm{R}}h^{ab} (\bm{R})  e^{-i\bm{k}\cdot  ( \bm{R}  + \bm{r}_a -\bm{r}_b)}$.
It is worth noting that the 
$h_{\bm{k}}^{ab}$ is \textit{not} 
periodic in the momentum space, 
i.e.,
$h_{\bm{k} +\bm{K}}^{ab} \neq h_{\bm{k}}^{ab} \label{eq:ham_trans}$
with $\bm{K} = 2\pi (n_x, n_y)$ ($n_{x,y} \in \mathbb{Z}$) 
being a reciprocal lattice vector. 
This is due to the fact that the sublattice positions, $\bm{r}_{a}$ and $\bm{r}_b$, are not the integer multiples of the lattice vectors.  
In the following, we use the notation that $h_{\bm{k}}$ denotes a $M \times M$ matrix 
whose $ab$-element is equal to $h_{\bm{k}}^{ab}$;
$M$ is the number of internal degrees of freedom.
Using the matrix representation of the Hamiltonian, one also obtain 
the (retarded) Green's 
function, $G_{\bm{k}}(\epsilon)= \left(
\epsilon + i\eta -h_{\bm{k}}
\right)^{-1}$, 
where $\eta$ corresponds to the damping rate. 

The velocity operator in the $\mu$-direction 
($\mu = x,y$) is given as
$v^{\mu} = \sum_{\bm{k},a,b}
[v^{\mu}_{\bm{k}}]^{ab}
c^{\dagger}_{\bm{k},a}c_{\bm{k},b}$
with
\begin{align}
v^{\mu}_{\bm{k}} = \frac{\partial h_{\bm{k}}}{\partial k_{\mu}}. \label{eq:v_def}
\end{align}
It is to be emphasized that the definition of the velocity operator in the form of 
Eq.~(\ref{eq:v_def}) is valid only when the Fourier transformation incorporates the sublattice position.

Now, let us consider the right-hand side of Eq.~(\ref{eq:GSvsKA}).
It is important to note that
the matrix $h_{\bm{k}}$ 
can be rewritten as 
\begin{align} 
h_{\bm{k}} = D_{\bm{k}} 
\bar{h}_{\bm{k} }
D^\dagger_{\bm{k}},
\label{eq:Unitary}
\end{align} 
where $D_{\bm{k}}$ is a unitary diagonal matrix
\begin{align}
[D_{\bm{k}}]_{ab} = e^{-i \bm{k} \cdot \bm{r}_a} \delta_{a,b}, \label{eq:def_D}
\end{align}
and $[\bar{h}_{\bm{k}}]^{ab} = 
\sum_{\bm{R}} h^{ab}(\bm{R}) e^{-i \bm{k}\cdot \bm{R}}$.
Note that $\bar{h}_{\bm{k}}$
satisfies
$\bar{h}_{\bm{k}+\bm{K}} = \bar{h}_{\bm{k}}$.
In the following, for clarity, the quantities denoted by $\bar{\cdot}$ have a periodicity in the momentum space
with respect to the translation by the reciprocal lattice vectors. 
Note that the Green's function is also written as
\begin{align}
G_{\bm{k}}(\epsilon)
= D_{\bm{k}} \bar{G}_{\bm{k}}(\epsilon)
D^\dagger_{\bm{k}}, \label{eq:green}
\end{align}
with $\bar{G}_{\bm{k}}(\epsilon) = (\epsilon+i\eta - \bar{h}_{\bm{k}})^{-1}$.
Further, 
by substituting Eq.~(\ref{eq:Unitary}) into Eq.~(\ref{eq:v_def}), 
we have 
\begin{align}
v^{\mu}_{\bm{k}} =
D_{\bm{k}} \bar{v}^{\mu}_{\bm{k}}D^\dagger_{\bm{k}}
+ \frac{\partial D_{\bm{k}}}{\partial k_\mu} \bar{h}_{\bm{k}}D_{\bm{k}}^\dagger
+ D_{\bm{k}} \bar{h}_{\bm{k}} \frac{\partial D_{\bm{k}}^\dagger}{\partial k_\mu}, \label{eq:v}
\end{align}
with 
$\bar{v}^{\mu}_{\bm{k}} = 
\frac{\partial \bar{h}_{\bm{k}}}{\partial k_{\mu}}$.
In fact, 
by explicitly performing the $k_{\mu}$ derivative of $D_{\bm{k}}$ and $D_{\bm{k}}^\dagger$, 
we find that Eq.~(\ref{eq:v}) can be rewritten as
\begin{align}
v^{\mu}_{\bm{k}} =
D_{\bm{k}} \left(
\bar{v}^{\mu}_{\bm{k}} + w^{\mu}_{\bm{k}}\right)D^{\dagger}_{\bm{k}},
\label{eq:velocity_2}
\end{align}
where
\begin{align}
w^{\mu}_{\bm{k}} = [\mathcal{R}_{\mu}, \bar{h}_{\bm{k}}], \label{eq:w}
\end{align}
with $\mathcal{R}_{\mu}$ being the diagnal matrix whose element is $[\mathcal{R}_{\mu}]_{ab} = -i r^{\mu}_a \delta_{a,b}$.
The key observation is that $w^{\mu}_{\bm{k}+\bm{K}} = w^{\mu}_{\bm{k}}$ holds,
since $\mathcal{R}_{\mu}$ is $\bm{k}$-independent and $\bar{h}_{\bm{k}}$ is periodic in the momentum space.
Hence, $w^{\mu}_{\bm{k}}$ is to be written as $\bar{w}^{\mu}_{\bm{k}}$ in our notation.
We further define $\bar{\gamma}^{\mu}_{\bm{k}}
= \bar{v}^{\mu}_{\bm{k}}+\bar{w}^{\mu}_{\bm{k}}$,
which is periodic in the momentum space.
Using $\bar{\gamma}^{\mu}_{\bm{k}}$, we can rewrite Eq.~(\ref{eq:velocity_2})
as 
\begin{align}
v^{\mu}_{\bm{k}} =
D_{\bm{k}} 
\bar{\gamma}^{\mu}_{\bm{k}}D^{\dagger}_{\bm{k}}.
\label{eq:velocity_3}
\end{align}

We now substitute Eqs.~(\ref{eq:green}) and (\ref{eq:velocity_3}) 
into the right-hand side of Eq.~(\ref{eq:GSvsKA}).
For the integrand of the first term, 
we have 
\begin{align}
&[\mathrm{Tr}\{ 
G_{\bm{k}}(\epsilon)v^y_{\bm{k}}
G_{\bm{k}}(\epsilon)v^x_{\bm{k}}
G_{\bm{k}}(\epsilon)v^y_{\bm{k}}
\} ]_{k_x=0}^{2\pi} \notag \\ 
= &[\mathrm{Tr}\{ 
(D_{\bm{k}} \bar{G}_{\bm{k}}(\epsilon)D^\dagger_{\bm{k}} )
(D_{\bm{k}}\bar{\gamma}^y_{\bm{k}}
D^\dagger_{\bm{k}} )
(D_{\bm{k}} \bar{G}_{\bm{k}}(\epsilon)D^\dagger_{\bm{k}}) \notag \\
& (D_{\bm{k}}\bar{\gamma}^x_{\bm{k}}
D^\dagger_{\bm{k}} )
(D_{\bm{k}} \bar{G}_{\bm{k}}(\epsilon)
D^\dagger_{\bm{k}})
(D_{\bm{k}}\bar{\gamma}^y_{\bm{k}}D^\dagger_{\bm{k}} )
\} ]_{k_x=0}^{2\pi} \notag \\ 
=& 
[\mathrm{Tr}\{ 
\bar{G}_{\bm{k}}(\epsilon)\bar{\gamma}_{\bm{k}}^y
\bar{G}_{\bm{k}}(\epsilon)\bar{\gamma}_{\bm{k}}^x
\bar{G}_{\bm{k}}(\epsilon)\bar{\gamma}_{\bm{k}}^y
\} ]_{k_x=0}^{2\pi} \notag \\
=& 0,\label{eq:conclusion}
\end{align}
since all the entries of the fourth line of Eq.~(\ref{eq:conclusion}) 
have the periodicity under $\bm{k} \rightarrow \bm{k}+\bm{K}$. 
The same argument holds for the second term 
of Eq.~(\ref{eq:GSvsKA}).
Therefore, we find that the right-hand side of Eq.~(\ref{eq:GSvsKA}) is vanishing, and thus $\chi^{\rm(KA)}_{\rm orb} =\chi^{\rm(GS)}_{\rm orb}$
exactly holds for the tight-binding models.

We remark that, using Eqs.~(\ref{eq:green}) and (\ref{eq:velocity_3}), 
as well as the similar relation for 
$\frac{\partial v^{\mu}_{\bm{k}}}{\partial k_{\nu}}$,
we can rewrite 
$\Theta^{(KA)/(GS)}(\epsilon)$ 
such that all the entries 
in the trace is periodic in the momentum space. 
However, 
it does not mean 
that these are 
the sublattice-position 
independent quantities.
Indeed, we can see that $\bar{w}^{\mu}_{\bm{k}}$ 
is sublattice-position-dependent 
because it contains $\mathcal{R}_\mu$ 
[Eq.~(\ref{eq:w})].
Note that the importance of 
the position dependence
has been recently emphasized by several literatures~\cite{Bena2009,Fruchart2014,Simon2020,Fuchs2021,Cayssol2021},
and it indeed affects some physical quantities. 

To summarize, we have proven that the Koshino-Ando's formula and the G\'{o}mez-Santos and Stauber's formula are eqiuvalent 
for the tight-binding models.
The difference between these two formulas is 
given by the surface term 
resulting from the partial integration. 
Importantly, the formal expressions of these formulas are valid only when 
a Fourier transform incorporating the sublattice position is used. 
Then, the Greens' functions and the velocity operators are not periodic 
in the momentum space,
which makes the treatment of the surface term somewhat delicate.
We have shown that the surface terms 
are exactly vanishing even in this case.

\begin{acknowledgement}
We thank 
Masao Ogata, Hideaki Maebashi, Hiroyasu Matsuura, Soshun Ozaki, and Masaki Kato for fruitful discussions.
This work is supported by JSPS KAKENHI
Grant No.~JP23K03243.
\end{acknowledgement}

\bibliographystyle{jpsj}
\bibliography{orb}

\begin{thebibliography}{10}
\bibitem{remark2}
: Our proof does not apply to cotinuum models where the Brillouin zone is
  absent.
\end{thebibliography}
\end{document}